\documentstyle{aipproc}

\begin{document}
\title{Radiation Processes in Blazars}
\author{Marek Sikora$^*$} 
\address{$^*$Nicolaus  Copernicus  Astronomical Center, PAN, 
Bartycka 18, 00-716 Warsaw, Poland}

\maketitle 
% Mathematical symbols \simgreat, \simless, \simprop:

\def\prop{\propto}
\newbox\grsign \setbox\grsign=\hbox{$>$} \newdimen\grdimen \grdimen=\ht\grsign
\newbox\simlessbox \newbox\simgreatbox \newbox\simpropbox
\setbox\simgreatbox=\hbox{\raise.5ex\hbox{$>$}\llap
     {\lower.5ex\hbox{$\sim$}}}\ht1=\grdimen\dp1=0pt
\setbox\simlessbox=\hbox{\raise.5ex\hbox{$<$}\llap
     {\lower.5ex\hbox{$\sim$}}}\ht2=\grdimen\dp2=0pt
\setbox\simpropbox=\hbox{\raise.5ex\hbox{$\prop$}\llap
     {\lower.5ex\hbox{$\sim$}}}\ht2=\grdimen\dp2=0pt
\def\simgreat{\mathrel{\copy\simgreatbox}}
\def\simless{\mathrel{\copy\simlessbox}}
\def\simprop{\mathrel{\copy\simpropbox}}

\def\doublespace{\baselineskip 24pt \lineskip 2pt \parskip 3pt plus 3pt}

\begin{abstract}
We  present an overview of the current theoretical 
models attempting to describe the structure and radiative processes 
operating in blazars  and discuss the observational constraints that these 
models must confront.  Many of these objects are found by {\it CGRO} to
be strong high energy $\gamma$--ray sources.  Their spectra consist of 
two broad components:  the low-energy component, peaking between the 
infrared and soft X--ray band, and the high-energy component, peaking 
in the high energy $\gamma$--rays.  The overall energetics is often 
dominated by the latter, by as much as an order of magnitude over the former.  
Both components show rapid large-amplitude variability, which indicates a 
very compact emission region.  

Most current models for the structure of blazars, to avoid the problems with 
excessive opacity of $\gamma$--rays to pair production, invoke 
beaming of electromagnetic emission from the radiating matter moving in 
a jet pointed close to the line of sight towards the observer, in an 
analogy to the jet-like structure inferred from other observations.
The non-thermal shape of the spectrum of the low-energy component as well 
as its polarization suggest synchrotron emission.  For the high-energy 
component, most current models invoke Comptonization of the lower energy 
photons, either those internal to the jet, as in synchrotron self-Compton 
(SSC) models, or external to the jet (UV radiation from the accretion disk or 
from the emission-line region, or IR radiation from dust), as in 
External Radiation Compton (ERC) models. Comparison of the energy densities 
of the external radiation fields with the energy density of 
the synchrotron radiation field  (both as measured in the jet comoving frame)
suggests that while the SSC model may be adequate to explain 
the $\gamma$--ray emission for BL Lac objects, the ERC  models are probably 
more applicable   for flat spectrum radio-quasars (FSRQ).

\end{abstract}

\section*{1. Introduction}

Blazars are extragalactic objects with core dominated, flat spectrum,
variable radio sources.  Radio spectra smoothly join the infrared-optical-UV 
spectra, and in all these bands flux is highly variable and  polarized. 
These properties are shared by BL Lac objects as well as by most flat spectrum
radio-quasars (FSRQ), and are successfully interpreted in terms of 
synchrotron radiation  produced in relativistic jets and beamed into our 
direction \cite{br78,bk79}. This interpretation is strongly
supported by direct observations of superluminal motions observed 
in radio cores in VLBI data \cite{veco94}, and lays the grounds for 
the unified scheme of radio-loud active galactic nuclei (AGN) 
\cite{gpcm93}. 

As was recently discovered by {\it CGRO}, many blazars are strong and 
variable sources of high energy $\gamma$--rays \cite{vonMo95};  in a 
few sources, the spectrum extends up to the TeV energies 
\cite{Punch92,Quinn96}.  
The $\gamma$--ray radiation  forms a separate spectral component, with
the luminosity peak located in the MeV-TeV range.
Variability of  GeV/TeV radiation itself provides  evidence for
relativistic speeds of radiating plasma \cite{matox93,dg95},
and the lack of
high energy $\gamma$--rays in AGNs other than blazars proves that
$\gamma$--rays must be at least as well collimated as
synchrotron radiation.

Production of high energy radiation  was predicted many years ago by
synchrotron-self-Compton (SSC) models \cite{kon81,mg85}. In this process, 
the same electrons that produce synchrotron radiation also upscatter some of 
synchrotron photons  to $\gamma$--ray energies.
However, as was recently recognized, the SSC process is not necesarily 
the one which produces most of the $\gamma$--rays.
The competing process, at least in quasars, may be the Comptonization 
of external radiation. Models based on this process are called  
ERC (external radiation Compton)  and their original 
variants have been investigated  by Dermer, Schlickeiser, \& Mastichiadis 
\cite{dsm92}, 
Sikora, Begelman, \& Rees \cite{sbr94}, Blandford \& Levinson \cite{bl95}, 
Ghisellini \& Madau \cite{gmad96}.
Production of $\gamma$--rays is also predicted by so-called hadronic models,
where ultrarelativistic electrons/positrons are injected by relativistic
protons \cite{manb92,man93,bed93,dl97}.

Different scenarios of $\gamma$--ray production in AGN jets are reviewed 
in \S 2.  Their predictions are confronted with multiwavelength spectral 
and variability data in \S 3, and discussion of what we have learned already 
and can learn in the nearest future about AGN jet physics from the blazar 
observations is presented in \S 4.  

\section*{2. Production of High Energy $\gamma$--rays}

The fact that the spectra of synchrotron components blazars extend up 
to optical, UV, and even X--ray bands indicates that AGN jets contain 
highly relativistic electrons/positrons, with Lorentz factors up to 
$10^4 - 10^6$.  Energy losses of such electrons are so rapid that they must be
accelerated/injected {\it in situ}, i.e. at the locations where they radiate.
These locations are the sites of the energy dissipation events, which 
propagate along the jet at moderate ($\Gamma \sim 10$) relativistic speeds.
The dissipation events can result from interaction of the jet with 
external obstacles, annihilation of magnetic fields, and/or collisions of 
inhomogeneities in a jet \cite{bk79}.  During such events, a part of the
dissipated energy is converted to relativistic electrons and protons.

Models based on the assumption that the high energy $\gamma$-rays 
are produced by directly accelerated electrons are discussed in \S 2.1,  
while hadronic models for $\gamma$--radiation production are  
reviewed briefly in \S 2.2.

\subsection*{2.1. ERC process vs. SSC process}

Assuming that the momentum distribution of relativistic electrons is isotropic
in the  comoving frame of dissipative events \cite{ds93}
and that
SSC and ERC operate  in the Thomson regime, one can compare
radiation production in these processes using formulae for
electron cooling rates 

\begin{equation}
\left (d\gamma' \over dt' \right )_{SSC} = 
{4 \over 3} {c \sigma_T \over m_e c^2} \,u_S' {\gamma'}^2
\end{equation}
and (\cite{sikora96})

\begin{equation}
\left (d\gamma' \over dt' \right )_{ERC} \simeq 
{16 \over 9} {c \sigma_T \over m_e c^2} \, u_{ext}' {\gamma'}^2
\end{equation}
where $\gamma'$ is the random Lorentz factor of relativistic electrons, 
\begin{equation}
u_S'= L_S'/4\pi a^2 c 
\end{equation}
is the energy density of 
synchrotron radiation produced by the source with radius $a$, 
\begin{equation}
u_{ext}' \simeq \Gamma^2 \xi L_{UV}/4 \pi r^2c
\end{equation}
is the energy density of the external radiation field \cite{sikora96}, 
$\xi$ is the fraction
of the central radiation isotropized by reprocessing and/or rescattering at
a distance scale corresponding with a distance $\sim r$  from
the central source, and  all primed 
quantities are as measured in the source comoving frame.

Using formulae (1) - (4) one can find that 
\begin{equation}
{L_{ERC}' \over L_{SSC}'} =
{(d\gamma'/dt)_{ERC} \over (d\gamma'/dt')_{SSC}} \simeq
(\Gamma \theta_j)^2 {\xi L_{UV} \over L_S'}
\end{equation}
where $\theta_j = a/r$. The above formula can be used to find 
the ratio of the observed SSC and ERC luminosities, provided 
that the angular distributions of radiation fields are known.
%Particularly, if SSC and ERC radiation fields  have the same angular
%distributions, then  
%$L_{ERC}/L_{SSC} = L_{ERC}'/L_{SSC}'$ is valid for any $\theta_{obs}$ 
%(which is true as long as these distributions are the same for the ERC 
%and SSC radiation).  
For isotropic (tangled) magnetic
field, the synchrotron radiation in the source  frame is isotropic, and 
for jets with $\theta_j \ll 1/\Gamma$, the observer located at $\theta_{obs}$ 
sees $L_S = {\cal D}^4 L_S'$, where 
${\cal D}= 1/(\Gamma (1-\beta \cos \theta_{obs}))$ is the Doppler factor.
As was pointed out by Dermer \cite{d95}, this
is not the case for the ERC process. Compton scattered radiation, as measured
in the source comoving frame, can have quasi-isotropic distribution only 
for $\Gamma \simeq \Gamma_{eq}$, where 
$\Gamma_{eq}$ is the Lorentz factor at
which the flux of the external radiation field is zero. 

As was shown by 
Sikora et al. \cite{sikora96} 
\begin{equation}
\Gamma_{eq} \simeq \left ( 3 \over 16 \xi \right )^{1/4} .
\end{equation}
At  $r \ll 1$ pc, $\xi$ is provided by rescattering
of central radiation by hot gas present in coronae or winds around 
accretion disks.  Up to $r \sim0.1 - 1.0 \sqrt {L_{UV}}$ pc, 
this is dominated by the fraction of the central radiation converted 
to emission lines by optically thick clouds/filaments, and at larger 
distances, it is determined by the fraction of the central radiation 
reemitted by dust.  For a typical quasar environment, $\xi$ is expected 
to be $\sim 0.01 - 0.1$, and, for such values, equation (6) gives 
$\Gamma_{eq} \le 2$. This is much less than the Lorentz factors of 
AGN jets which are typically enclosed in the range $5 < \Gamma < 20$
and often
reach values $>10$ \cite{veco94,padur92,mr94}.
The external radiation is seen by the emitting cloud of plasma moving at 
$\Gamma > \Gamma_{eq}$ as coming from the front. 
In such a field, the Compton scattered radiation is produced with 
a highly anisotropic distribution, with a
large deficiency of radiation scattered into the $\theta' > \pi/2$-hemisphere.
As measured in the external frame, such radiation is beamed more strongly 
than the synchrotron radiation and the  observed luminosity $L_{ERC} \propto
{\cal D}^6$.  However, for
comparable total emitted luminosities, 
the respective observed luminosities $L_{SSC}$ and $L_{ERC}$, if averaged 
over the $1/\Gamma$-cone, are also comparable. 
Then, for the case $\theta_j \sim 1/\Gamma$, for which the angular
distribution of radiation  within the $1/\Gamma$-cone is smeared out,
the formula
\begin{equation}
L_{ERC}/L_{SSC} \sim L_{ERC}'/L_{SSC}' \sim 
(\Gamma \theta_j)^2 {\xi L_{UV} \over L_S} \Gamma^4
\end{equation} 
can be used  for all $\theta_{obs} \le 1/\Gamma$-observers.
Since in  quasars $\Gamma \sim 10$ and
$\xi L_{UV} \sim 10^{44} - 10^{45}$ erg s$^{-1}$, while $L_S$ observed in 
FSRQ is in the range $10^{46} - 10^{47}$ erg s$^{-1}$, the above formula 
seems to prove strong domination of the ERC process over the SSC process 
in quasar jets. 

However, it should be noted here that because of the different energy 
distributions of the ambient radiation fields, 
the respective high energy components produced by a given
population of electrons will not overlap entirely. As a result, a less 
luminous SSC component produced by far infrared radiation can still be 
visible in soft/mid X--rays \cite{kubo97}, while relativistic electrons 
injected  at a distance $\sim 1$pc can produce in the  MeV range two separate 
``bumps'', one due to Comptonization of external UV radiation and one due to
Comptonization of external near-IR radiation. 

The situation is less clear in BL Lac objects, where the radiative environment
in the central region is not very well known. The lack of strong emission lines
and of UV excesses (even during lowest states) suggest that in these
objects $\xi L_{UV}$ can be very low. Noting also that in BL Lac objects 
$\Gamma$ factors are typically smaller than in quasars \cite{padur92,mr94}, 
domination of SSC over ERC in these objects is very likely.  
%This is 
%further supported by a good agreement of the values of Lorentz factors of the 
%radiating electrons that are required to produce the TeV radiation as 
%compared to the values inferred from spectral variability observed 
%in X--rays \cite{ taka96}.  

One should be warned, however, that for BL Lac objects with the high energy 
spectra extending  up to TeV energies the formula (7) cannot be applied 
directly;  this is because the Klein-Nishina effect reduces the efficiency 
of the Compton process, and this reduction is different for different energy 
distributions of the ambient radiation fields. 

\subsection*{2.2. Hadronic Models}

In all particle acceleration processes, the injection of relativistic 
electrons/positrons is accompanied by injection of relativistic protons.
Their energy can be converted to high energy radiation following such 
processes as direct synchrotron radiation of protons, 
proton-photon pair production, photomeson production, and nuclear collisions. 
The first three processes are known to be very inefficient, and in AGN jets
can become important only for proton energies $\ge 10^8 - 10^{10}$ GeV. 
Only for such high energies can the time scales of the proton energy
losses  become comparable to or shorter than the propagation time scale of 
the source in a jet.  Energy losses of such energetic protons are dominated 
by  photomeson production, and this process was used by Mannheim and 
Bierman \cite{manb92} to model $\gamma$--ray production in luminous blazars.

The radiation target for  photomeson production is dominated by the 
near/mid-infrared radiation. In quasars, such radiation is provided by hot dust
at distances $\sim 1-10$ parsecs from the central source and by synchrotron
radiation in a jet, produced by directly accelerated electrons.
The main output of the photomeson process are single pions. They take about
 30\% of the protons' energy and convert it to photons, neutrinos, and 
through muons, to electrons and positrons.
The photons injected by neutral pions are immediately absorbed by 
soft photons in the pair production process. These pairs 
and  electrons/positrons injected by 
muons have Lorentz factors $\gamma' \ge 10^{11}$. For such energies, Compton
scattering with the ambient radiation field takes place deeply in 
the Klein-Nishina
regime and, therefore, their energy losses are dominated by
synchrotron radiation. Most of this radiation is so energetic that it 
produces two more generations of photons and pairs. The final output of this
synchrotron-supported pair cascade is the high energy component, enclosed
within or cut off at energies above which photons are absorbed by 
$\gamma\gamma$-pair production process.
This maximum energy can be $\sim 30$ GeV in FSRQ, as determined by
external UV radiation, and $\sim 1$ TeV in low luminosity BL Lac objects, as
determined by infrared radiation of dust \cite{pb97}.

The weakness of the ``photomeson'' model is that it requires fine tuning
in order to avoid situations where the luminosity peak lies below MeV energies.
This is because, after 3 pair generations, the  location of the peak depends
on the 6th power of the maximum proton energy. Also, even if a model is 
successful in locating the peak of the high energy component, there is 
still the  problem of how to obtain  the hard X--ray spectra after 
three generations 
of the pair cascade process \cite{sve87}. To overcome this difficulty, 
Mannheim \cite{man93}
proposed that transition from softer $\gamma$--ray spectra to harder 
X--ray spectra results from a break in the pair injection function.
This, however, requires the ambient radiation to be transparent to 
$\gamma$--rays up to energies $\sim 10 \sqrt {\Gamma/B'} $ TeV, while  
external 
UV and near-IR radiation fields are expected to cut the spectrum in 
quasars at $\sim 30$ GeV for $r < \sqrt {L_{UV}}$ pc and
at $\sim 1$ TeV for larger distances.

The photomeson scenario was also suggested to explain the production 
of TeV radiation in low luminosity BL Lac objects \cite{pb97}. The recent 
discovery of  variability on time scales $< 1$ hour seems to jeopardize
this idea. This is because to get proton energy losses on such short time 
scales, much higher IR luminosities are required than are observed.

Much less extreme  proton energies are required in models based 
on the assumption that 
proton energy losses are dominated by collisions with the ambient gas.
The final output of these  collisions is the same  as in the photomeson
process, i.e.,  relativistic electrons/positrons, photons and neutrinos.
The process can be efficient only if the column density of 
the target is $n_H \ge 10^{26}$ cm$^{-2}$.  Bednarek \cite{bed93} proposed 
as a target the funnels formed around the black hole
by a geometrically thick disk, while Dar and Laor \cite{dl97} 
suggested interactions of jet with clouds and/or stellar winds. 
The shortcoming of such models is that relativistic protons, 
before colliding with the nuclei, may easily 
suffer deflections by magnetic fields; 
  this generally results in a lack of 
collimation of the radiation produced following pp collisions.  

\section*{3. Multiwavelength Spectra}

\subsection*{3.1. ERC and SSC luminosities vs Synchrotron Luminosity}

If both high-energy and low-energy spectral components are produced 
by the same population of relativistic electrons, and the production of 
high energy radiation is dominated by the SSC process, then 
\begin{equation} 
{L_{SSC} \over L_S} \sim {L_{SSC}' \over L_S'} \sim {u_S' \over u_B'}
\end {equation}
where  $u_B'= (B')^2/8\pi$ is the energy density of magnetic field. 
Equations~(3) and (8) give 
\begin{equation} 
u_B' \simeq {1 \over 4 \pi a^2 c \Gamma^4}\, {L_S^2 \over L_{SSC} },
\end{equation}
which, in the case of steady flow with $\theta_j \sim a /r$, determines 
the flux of magnetic energy
\begin{equation}
L_B \simeq c u_B' \pi a^2 \Gamma^2 = 
{1 \over 4 \Gamma^2} {L_S^2 \over L_{SSC}} .
\end{equation}
For $\Gamma \sim 10$, $L_S \sim 10^{46} - 10^{47} {\rm erg s}^{-1}$ 
this gives $L_B \sim 10^{43} {\rm erg s}^{-1}$, which is about
3 orders of magnitude less than the typical power of the quasar jets 
\cite{rawsa91,cf93,fmb95}.
Thus, the observed high $\gamma$--ray luminosities can be explained  
in terms of the SSC models only if one assumes that the jets are very weakly 
magnetized.
%(\cite{rawsa91}). 

In the case of ERC models we have 
\begin{equation} 
{L_{ERC}' \over L_S'} \sim {u_{ext}' \over u_B'} , 
\end{equation} 
and provided that $L_{ERC}$ and $L_S$ are  luminosities observed 
at $\theta_{obs} \le 1/\Gamma$ and that $\theta_j \sim 1/\Gamma$, we can 
use the scaling 
$L_{ERC}/L_S \sim L_{ERC}'/L_S$. With this scaling equations (4) and  (11)
give 
\begin{equation} 
u_B' \simeq \Gamma^2 \, { \xi L_{UV} \over 4 \pi r^2 c}\, 
{L_S \over L_{ERC}}. 
\end{equation}
This gives
\begin{equation}
L_B \simeq c u_B'\pi a^2 \Gamma^2 = 
{(\theta_j \Gamma)^2  \over 4} \Gamma^2 \xi L_{UV} {L_S \over L_{ERC}} 
\end{equation}
which is 2-3 orders of magnitude greater than in the case 
of the SSC model. 

\smallskip
Another interesting aspect of comparing ERC and SSC radiation 
components with the synchrotron component is the 
angular distribution of these radiation fields.  As was shown by Dermer 
\cite{d95} and discussed in \S 2.1, ERC radiation is much more strongly 
collimated than the synchrotron and SSC radiation. Since SSC and synchrotron
radiation fields have the same angular distribution (they both are
produced isotropically in the source comoving frame), the predicted 
high-energy to low-energy luminosity ratio doesn't depend on $\theta_{obs}$.
In contrast, the ERC model predicts this ratio to drop very rapidly with
viewing angle outside the $1/\Gamma$-cone.
Dermer proposed that this can explain why a significant fraction of FSRQ 
do not show $\gamma$--ray activity, even though they are 
otherwise recognized as typical blazars
on the basis of the low energy component properties.

\subsection*{3.2. Production of  X--rays}

X--rays in different sub-classes of blazars can have different origins. 
In most BL Lac objects X--ray spectra are steep ($\alpha \sim 1 - 3$) 
and variable, and lie on an extrapolation of the UV spectrum.  This 
indicates that X--rays in these objects represent high energy tails of the 
synchrotron component.  In FSRQs, the X--ray spectra are usually very hard 
($\alpha \simeq 0.5 - 0.7$), showing weaker variability than in other 
spectral bands. These spectra are often interpreted as low energy tails 
of the $\gamma$--ray components;  however, one cannot exclude the 
possibility that they are superposed from two or more components.
And finally, there are intermediate objects where the soft X--rays are 
dominated (at least occasionally) by the synchrotron component, while
higher energy  X--rays belong to the high-energy Compton component 
\cite{mad96}. These differences in the X--ray spectra 
seem to follow the general trend where in  less luminous 
blazar, the peaks 
of the low-energy (synchrotron) component are located at higher energies 
\cite{smu96,kubo97}.  

%\subsubsection*{3.2.1. More about X--rays in FSRQ}   

The simplest interpretation of the hard X--ray spectra of FSRQs is that, 
together with the $\gamma$--rays, they form a single component produced by the 
ERC process.  In the model, the spectral slope changes from a steeper one 
in the $\gamma$--ray band to a harder one in the X--ray band, as a result 
of incomplete cooling of electrons  below certain energy which
is determined  by an equality of the ERC cooling time scale and the  
propagation time scale \cite{sbr94}.
The break is located in the $1 - 30$ MeV range if the distance of radiation
production is $\sim 0.1 - 3$ pc.

In ERC  models, the X--ray spectra imprint the distribution of Lorentz 
factors of relativistic electrons down to $\gamma' \sim 10$ for the 
X--rays produced by Comptonization of IR radiation, and even down to 
$\gamma' \sim 1$ for the X--rays produced by Comptonization of UV radiation.
Since for distances $> 0.1$ pc the low energy electrons cool 
very inefficiently, to produce the observed X--ray flux 
by ERC process requires 
such a large number of electrons that the jet must be strongly 
pair dominated in order to avoid unreasonable high kinetic energy flux.

%The MeV-break can also result from the break in the pair injection function,
%in models where pair cascades are involved. 
%In the pair cascade determined  by ERC radiation  and by absorption of
%$\gamma$--rays by external X--rays, the break at MeV energies 
%corresponds with the pair injection cutoff at 100 MeV energies. In this 
%scenario the $\gamma$--ray spectra  above 100 MeV result
%from superposition of radiation produced by pair cascades over 2
%distance decades, starting from the distance $\sim 10^{16}$ cm at which
%the MeV break is formed.

Another possibility  is that hard X--ray spectra are superposed from
partial spectra produced over a wide range of  distances and having
low-energy cutoffs at energies which increase with  distance
\cite{smmp97}. In this model, the soft X--rays are produced closely 
to the black hole, and therefore the production of X--rays can be 
accomplished 
by a lower number of electrons, and thus the jet plasma need not be 
pair-dominated.  

Finally, the hard X--ray spectra can be produced by SSC radiation,
while production of high energy $\gamma$--rays can be dominated by ERC process
\cite{kubo97}. In this model, a wide range of $n_e/n_p$ is acceptable.

\subsection*{3.3. Bulk-Compton Radiation}

Very interesting constraints on the AGN jets come not only from what we 
{\it do} observe, but also from what we {\it do not}  observe. A feature 
that was predicted -- but not confirmed observationally -- 
is  radiation produced by cold electrons in a jet.
Such electrons, dragged by the protons and/or magnetic fields in the jet, 
for $\Gamma > \Gamma_{eq}$, should  scatter external UV photons and produce
a collimated beam of bulk-Compton radiation \cite{bs87,sikora96}.
The predicted observed luminosity  is 
\begin{equation} 
L_{BC} \sim \Gamma^2 n_e \pi a^2 r dE_e/dt
\end{equation}
where
\begin{equation}
{dE_e \over dt} = {4\over 3} c \sigma_T \xi u_{ext} \Gamma^2,
\end{equation}
$u_{ext} \sim L_{UV}/4 \pi r^2 c $, and $r$ is the distance at which
this process is most efficient.
The bulk-Compton spectrum should have a peak at
$h\nu_{BC} \sim \Gamma^2 n_{UV} \sim 1{\rm keV}$.  Since this is 
not observed, $L_{BC}$ must 
be smaller than the luminosity $L_{SX}$ of the nonthermal X--ray spectrum 
in the soft X--ray band.  This gives an upper limit for 
the Thompson optical thickness in a jet
\begin{equation}
\tau_{j,max} \equiv  n_{e,max} a \sigma_T \sim 
{3 \over \Gamma^3 (\theta_j \Gamma)}\, {L_{SX} \over \xi L_{UV}} 
\end{equation}
i.e., $\sim 0.03$ for $\Gamma \sim 10$, $L_{SX} \sim 10^{46}$ erg s$^{-1}$,
 and  $\xi L_{UV} \sim 10^{45}$ erg s$^{-1}$, 
With this limit, the  processes scaled by
$(\tau_j)^2$ (like annihilation, bremsstrahlung and Coulomb interactions) 
are inefficient, and play a negligible role in shaping the spectra of blazars. 
This is because in such thin plasmas, the time scales of these processes 
are much longer than the time scale of plasma propagation in a jet \cite{cb90}.

The upper limit for $\tau_j$ also gives interesting constraints on 
the $e^+e^-$ pair content of a jet. If $r_{min}$ is the radius where $\tau_j$ 
is maximal 
and if for $r>r_{min}$ the pair flux is conserved, then for a conical 
jet $\tau_j \propto 1/r$ and, therefore, the bulk-Compton radiation is mostly 
contributed by the innermost parts of the jet. Assuming that the 
kinetic energy flux in a jet is dominated by cold protons, we have  
\begin{equation}
L_K  \simeq n_p' m_pc^3 \pi a^2 \Gamma^2 
\end{equation}
where $n_p'$ is the number density of protons in the jet comoving frame.  
Noting that $n_p = n_p' \Gamma$, we have
\begin{equation} 
n_e \simeq 
{n_e \over n_p} {L_{K} \Gamma \over \pi m_p c^3 a^2 \Gamma^2}. 
\end{equation} 
Substituting this into the formula for $L_{BC}$ evaluated for $r = r_{min}$, 
we obtain
\begin{equation} 
 L_{BC} \simeq
{n_e \over n_p} {r_g \over r_{min}}\, {L_K \over L_{Edd}}\, \xi L_{UV} 
\Gamma^3,  
\end{equation}
where $L_{Edd}= (4 \pi m_p c^3 /\sigma_T) r_g $.
Then, the condition $L_{BC} \le L_{SX}$ gives that for 
powerful ($L_K \sim L_{Edd}$) and  pair-dominated 
($n_e \gg n_p$) jets, overproduction of soft X--rays can be  
avoided only at very large distances ($10^3 - 10^5 r_g$) from the black hole. 

\section*{4. Summary}

Discovery of strong and variable $\gamma$--ray radiation in blazars by CGRO
provided an exceptional possibilities to explore and verify the operation 
of various nonthermal processes in AGN jets, and to study the structure, 
energetics and matter content of these jets. The main achievements of such 
studies and future prospects are listed below:
\smallskip

$\bullet$ $\gamma$--ray radiation provides independent evidence that 
blazar radiation is produced by relativistic jets \cite{matox93,dg95}. 
This is because the compactness of the source derived from the observed 
$\gamma$- and X--ray luminosities and variability time scales is so high 
that if it was  intrinsic (true) compactness, all $\gamma$--rays would be 
absorbed  by $\gamma\gamma$ pair production process.
This implies that the true source compactness must be much lower than 
the observed one, and this is the case  if the observed radiation 
originates from  plasma propagating in our direction at  relativistic 
speed.
\smallskip

$\bullet$ $\gamma$--rays can be also absorbed by  external radiation fields,
and because the  compactness of such fields decreases with  distance,
this gives the  minimum distance from which  the $\gamma$--rays can escape.
Of course, since the opacity for $\gamma\gamma$ interactions depends on 
energy and is higher for more energetic $\gamma$--rays, the minimum escape
distance is smaller for less energetic $\gamma$--rays.
 Such stratified 
$\gamma$--ray production was suggested by Blandford and Levinson 
\cite{bl95} in their inhomogeneous version of ERC model.

\smallskip

$\bullet$ Huge apparent $\gamma$--ray luminosities, reaching in some FSRQ  
$10^{48}-10^{49}$ erg s$^{-1}$,  provide independent
evidence  that AGN jets must be very powerful, with
$P > (\bar L_{\gamma}/\Gamma^2)/\epsilon_{rad}$, i.e., 
$ \sim  10^{46}$ erg s$^{-1}$ for radiation efficiency 
$\epsilon_{rad} \sim 0.1$, 
$\Gamma \sim 10$ and time-averaged observed luminosity $\bar L_{\gamma}
\sim 10^{47}$ erg s$^{-1}$. 
\smallskip
 
$\bullet$ Comparison of the ERC efficiency with the SSC efficiency for 
typical quasar radiation fields implies that the production of $\gamma$--rays 
in FSRQ should be strongly dominated by Comptonization of external radiation. 
However, SSC can still contribute visibly to the X--ray band \cite{kubo97}.
\smallskip

$\bullet$ The hadronic models have problems explaining hard  X--ray spectra 
in FSRQ and very short variability time scales in TeV BL Lac objects.
\smallskip

$\bullet$ Comparing ERC and SSC spectra with the synchrotron spectra, 
one can attempt to  derive  physical parameters of radiating plasma - such as 
maximum electron energies, magnetic fields, electron injection function, and 
 distance of the source from the black hole 
\cite{sbr94,gmd96,smmp97,taka96,mk97}. 
However, such analyses must be performed by taking into account that the 
observed spectra, especially their lower energy parts (in both the 
synchrotron and the Compton components) may well be superposed by
two or more components.
\smallskip

$\bullet$ 
In the case of FSRQ, simultaneous observations in the $\gamma$--ray
and X--rays bands must be used to verify various mechanisms of X--ray 
production. This can help to establish the pair content in AGN jets.
In particular, the correlation and absence of a time lag between 
the $\gamma$--ray 
and the X--ray flares (as seen in Jan, 1996 in 3C279;  \cite{Weh97}) 
may indicate co-spatial production of both, and in the case of 
X--rays produced by the ERC process, this would indicate 
a pair-dominated plasma.

\smallskip

$\bullet$.
Soft X--ray limits for the bulk-Compton process prove that  AGN jets are
optically thin. This implies that such processes as bremsstrahlung, 
annihilation and Coulomb interactions are inefficient in these jets.  

\smallskip

$\bullet$
For a given $n_e/n_p$, the upper limit for the bulk-Compton radiation
gives a minimum distance for jet formation.  For strongly 
pair-dominated plasmas this distance is $0.1 - 1.0$ pc.  However,
jets with energy flux $P \ge 10^{46}$ erg s$^{-1}$ must be  powered 
very near the black hole, by its rotation \cite{bz77} or by the 
innermost parts of an accretion disk \cite{bp82}.  These two inferences 
-- the large distances of formation of pair-dominated jets
and the very central source of the jet energy -- can be reconciled
if over the first 3 decades of distance, the jet is strongly dominated by 
the Poynting flux. This very wide distance range of conversion 
of the magnetic energy to the bulk kinetic energy can result from 
radiation drag \cite{lbc92}.  This transition process can be accompanied 
by pair production in shocks and magnetic field reconnection sites
\cite{l96,rolo97}.

\bigskip
\leftline {\bf ACKNOWLEDGMENTS}
I wish to thank Mitch Begelman and Greg Madejski for  valuable comments which
helped improving the paper. This work has been supported in part by the NASA
grant NAG5-4106 and by the Polish KBN grant 2P03D01209.

\end{document}